\documentclass[12pt]{article}
\usepackage{graphicx}
\usepackage{amssymb}
\usepackage{amsfonts}

\title{Quantum entanglement: an overview of the separability problem in two quantum bits }
\author{Honorine Gnonfin$^{1, \star}$ and Laure Gouba$^{2,3,\diamond}$ \\
$^{1}${\em
African Institute for Mathematical Science (AIMS-S\'en\'egal)} \\
{\em Km 2 routes de Joal (Centre IRD Mbour) B. P. 1418 Mbour, S\'en\'egal.}\\
$^{\star}${\em Email: honorine.gnonfin@aims-senegal.org}\\
$^{2}${\em
The Abdus Salam International Centre for Theoretical Physics (ICTP)}\\
{\em Strada Costiera 11, I-34151 Trieste Italy.}\\
$^{3}${\em International Chair of Mathematical Physics
 and Applications (ICMPA)} \\
 {\em UNESCO Chair, University of Abomey-Calavi (UAC)}\\
 {\em 072 B.P. 50 Cotonou, Republic of Benin.}\\
$^{\diamond}$ {\em Email: laure.gouba@gmail.com}}

\begin{document}

\maketitle

\begin{abstract}
\noindent
The separability problem is one of the basic and emergent problems in present and future quantum information processing. The latter focuses on information and computing based on quantum mechanics and uses quantum bits as its basic information units. In this paper we present an overview of the progress in the separability problem in bipartite systems, more specifically in two quantum bits (qubits) system, from the criterion based on the Bell's inequalities in 1964 to the Li-Qiao criterion and the enhanced entanglement criterion based on the SIC POVMs in 2018.
\end{abstract}

\section{Introduction}

Quantum mechanics is a beautiful and fascinating theory, developed in fits, and started from $1900$ to $1920s$, maturing into its present form in the 
late $1920$s. A collection of views about the meaning of quantum mechanics principally attributed to Niels Bohr and Werner Heisenberg is called the Copenhagen interpretation \cite{copen1}. There is no definitive historical statement of what the Copenhagen interpretation is. It is one of the oldest and numerous proposed interpretations of quantum mechanics as features of it date to the development of quantum mechanics during $1925$ - $1927$, and it remains one of the most commonly taught \cite{copen2}.
Albert Einstein was skeptical of quantum mechanics, particularly its copenhagen interpretation \cite{epr1}. In the May 15 $1935$ issue of Physical Review, Albert Einstein co-authored a paper with Boris Podolsky and Nathan Rosen who were his two postdoctoral research associates at the Institute for Advanced Study. The article was entitled {\it Can quantum mechanical description of physical reality be considered complete}? \cite{epr2}.
In this study, the three scientists proposed a thought experiment known today as EPR paradox that attempted to show that the quantum mechanical description of physical reality given by wave functions is not complete.

However, Einstein, Podolsky and Rosen did not coin the word entanglement. Erwin Schr\"odinger in his correspondence with Einstein, following the EPR paper used the word Verschr\"anking (in German) translated by himself in English as entanglement, to describe the correlations between two particles that interact and then separate as in the EPR thought experiment.
He shortly thereafter published a seminal paper defining and discussing the notion of entanglement \cite{schr}. In this seminal paper, Schr\"odinger recognized the importance of the concept, and stated : `` I would not call (entanglement) one but rather the characteristic trait of quantum mechanics, the one that enforces its entire departure from classical lines of thought ". Einstein was disturbed by the theoretical concept of quantum entanglement, which he called {\it Spooky action at distance}. Einstein did not believe two particles could remain connected to each other over great distances: doing so, he said, would require them to communicate faster than the speed of light, something he had previously shown to be impossible.
Like Einstein, Schr\"odinger was dissastified with the concept of entanglement, because it seemed to violate the speed limit on the transmission of information implicit in the theory of relativity. 
The EPR paper generated significant interest among physicists and quickly became a centerpiece in debates over the interpretation of quantum theory, debates that continue today.

The question of expected locality of the entangled quantum systems raised by EPR allowed John Stewart Bell to discover his famous inequalities serving as a test and demonstration of strange properties of the simplest entangled wave function represented by a singlet state \cite{Bell1, Bell2}.  Still one had to wait long for the proposals and pratical applications of quantum entanglement. 
Until $1975$, a decisive experiment based on the violation of Bell's inequalities and verifying the veracity of quantum entanglement was missing. The experiment led by French physicist Alain Aspect at the Ecole Sup\'erieure d'optique in Orsay between $1980$ and $1982$ was the first quantum mechanics experiment to demonstrate the violation of Bell's inequalities \cite{aspect1, 
aspect2}. This experiment is called the Aspect's experiment. It confirmed the predictions of quantum mechanics and thus confirmed its incompabilities with local theories. Quantum entanglement is a phenomenon which has no counterpart in classical physics. It can be seen as the most non-classical feature of quantum mechanics that has risen numerous philosophical, physical and mathematical questions since the early days of the quantum theory. 
A paper by D. Bruss and al., published in 2001 listed four important issues of motivation to study the entanglement problem \cite{bruss1}:
\begin{enumerate}
\item The essential role of entanglement in apparent ``paradoxes" and counter-intuitive consequences of quantum mechanics \cite{epr2, schr, wheeler}.
\item The characterization of entanglement is one of the most fundamental open problems of quantum mechanics \cite{peres}.
\item Entanglement plays an essential role in applications of quantum mechanics to quantum information processing \cite{bouw, nielchuan},  quantum cryptography \cite{bruss2, giriti}, and quantum communication \cite{bennett}.
\item The entanglement problem is directly related to one of the most challenging open problems of linear algebra and functional analysis: the characterization and classification of positive maps on $C^\star$ algebras \cite{stro, jami, choi, woro}.
\end{enumerate}

 A recent paper by P. Horodecki and al., titled {\it Five Open Problems in Quantum Information Theory} includes open problems in quantum metrology, quantum entanglement and its distillability \cite{horodecki2}. This paper boosts our motivation and deepens our interest in studying quantum entanglement. The definition of entanglement is nowadays a mathematical one, that is rather simple, compare to the phenomenological description that is still difficult.

The wave function describing quantum system is entangled if it cannot be
written as a product states of subsystems. The simplest example is the singlet state of two spin-$\frac{1}{2}$ particles \cite{bohm}
\begin{equation}
|\psi\rangle = \frac{1}{\sqrt{2}}\left( |0\rangle|1\rangle - |1\rangle|0\rangle \right).
\end{equation}
It can be proven that $|\psi\rangle \neq |\varphi\rangle |\phi\rangle$ for any $|\varphi\rangle$ and $|\phi\rangle$ describing subsystems, $|0\rangle$ standing for ``spin up" state and $|1\rangle$ standing for ``spin down" state. 
The `` nonfactorisability" of any bipartite pure state implies that its reduced density matrices are mixed. The above definition is naturally generalized to the entanglement of multiparticle pure state. A bipartite pure quantum state $|\psi\rangle_{AB}\in \mathcal{H}_A \otimes \mathcal{H}_B$ is called entangled when it cannot be written as 
\begin{equation}
|\psi\rangle_{AB} = | \psi\rangle_A \otimes |\psi\rangle_B,
\end{equation} 
 for some $|\psi\rangle_A \in \mathcal{H}_A$ and $|\psi\rangle_B \in \mathcal{H}_B$. A mixed state or density matrix $\rho_{AB}$ which is semi-definite operator on $\mathcal{H}_A \otimes \mathcal{H}_B$ is called entangled when it cannot be written in the following form 
\begin{equation}
 \rho = \sum_i p_i|\psi_i\rangle_A{}_A\langle \psi_i| \otimes |\psi_i\rangle_B{}_B\langle \psi_i|
\end{equation}
 Here the coefficients $p_i$ are probabilities, that means $0\le p_i \le 1$ and $\sum_{i} P_i = 1 $. Note that in general neither $\{|\psi_i\rangle_A\}$
 nor $\{|\psi_i\rangle_B\}$ have to be orthogonal.

The fundamental question in quantum entanglement theory is which states are entangled and which states are not and this question is still an open problem today, both from the theoretical and experimental point of view, and  known as the separability problem, that has been solved for pure states \cite{neven}, and for $2 \times 2$ and $2 \times 3$ systems \cite{horo}. A separability condition can be necessary or necessary and sufficient conditions for separability. A necessary condition for separability has to be fulfilled by every separable state. In that case, if a state does not fulfill the condition, it has to be entangled, but if it fulfills we cannot conclude.
On the other hand, a necessary and sufficient condition for separability can only be satisfied by separable states, if a state fulfills a necessary and sufficient condition, then we can be sure that the state is separable \cite{horofami}.

In this paper, we attempt to present an overview on the operational and non operational criteria of separability in two quantum bits (qubits).
The manuscript is organized as follows. In section (\ref{sec2}), we present the operational criteria followed by the non operational criteria in section (\ref{sec3}). The conclusion is given in section (\ref{sec4}).

\section{Operational criteria}\label{sec2}

An operational criterion is a recipe that can be applied to an explicit density matrix $\rho$, giving some immediate answer like `` $\rho$ is entangled" or ``$\rho$ is separable" or this ``criterion is not strong enough to decide whether $\rho$ is separable or entangled ".
\begin{enumerate}

\item {\it Bell-CHSH inequalities}

The Bell inequality was originally designated to test predictions of quantum mechanics against those of a local hidden variables theory \cite{Bell1}. 
Bell's inequalities were initially dealing with two qubits, i.e two-level systems and provide a necessary criterion for the separability of 2-qubits states. For pure states, Bell's inequalities are also sufficient for separability.  It has been proven by Gisin that any non-product state of two-particle systems violates a Bell-inequality \cite{gisin}. 
This inequality which involves three vectors in real space $\mathbb{R}^3$ determing which component of a spin to be measured by each party or three, has been extended for the case involving four vectors by Clauser, Horne, Shimony and Holt (CHSH) in 1969 \cite{chshin}. The Bell-CHSH inequality also provides a test to distinguish entangled from non-entangled states.

Consider a system  of two qubits. Let $A$ and $A'$ denote observables on the first qubit, $B$ and $B'$ denote observables on the second qubit, the Bell-CHSH inequality says that for non-entangled states, means for states of the form $\rho = \rho_1\otimes \rho_2$, or mixtures of such states, the following inequality holds:
\begin{equation}\label{bin1}
| \langle\; A\otimes B + A\otimes B' + A'\otimes B - A'\otimes B'\;\rangle_\rho|\le 2,
\end{equation}
where $\langle A\otimes B \rangle_{\rho} := \textrm{Tr}\; \rho (A\otimes B)$ and 
$\langle A\otimes B \rangle_{\psi} = \langle \psi |A\otimes B|\psi\rangle $
for the expectation value of $A \otimes B$ in the mixed states $\rho$ or pure state $|\psi\rangle$.
As an example, we consider a two qubits state $|\phi\rangle = \frac{1}{\sqrt{2}} (|00\rangle + |11\rangle)$, and the observables 
\begin{equation}
A = \frac{1}{\sqrt{2}}(\sigma_x +\sigma_z),\: A' = \frac{1}{\sqrt{2}}(\sigma_x - \sigma_z), \: B = \sigma_x, \: B' = \sigma_z,
\end{equation}
where $\sigma_x$ and $\sigma_z$  are Pauli matrices. We have then explicitly
\begin{equation}
\sigma_x = \left(\begin{array}{cc}
0 & 1\\
1 & 0
\end{array} \right); \quad \sigma_z = \left(\begin{array}{cc}
1  & 0\\
0 & -1
\end{array} \right); 
\end{equation}
and
\begin{equation}
A  = \frac{1}{\sqrt 2}\left(\begin{array}{cc}
1 & 1\\
1 & -1
\end{array} \right);\: A' = \frac{1}{\sqrt 2}\left(\begin{array}{cc}
-1 & 1\\
1 & 1
\end{array} \right);\: B = \left(\begin{array}{cc}
0 & 1\\
1 & 0
\end{array} \right); \: B' = \left(\begin{array}{cc}
1 & 0\\
0 & -1
\end{array} \right).
\end{equation}
It is easy to check that 
\begin{equation}
\langle \phi|\; A\otimes B + A\otimes B' + A'\otimes B - A'\otimes B'\;|\phi\rangle = 2\sqrt 2.
\end{equation}
The state $|\phi\rangle$ which violates the Bell-CHSH inequality is a well known entangled state, and is one of the Bell pairs, maximally entangled state.
The maximal violation of (\ref{bin1}), for entangled states follows from an inequality of Cirelson \cite{cirelson}
\begin{equation}\label{bin2}
| \langle A\otimes B + A\otimes B' + A'\otimes B - A'\otimes B'\rangle_\rho|\le 2\sqrt 2.
\end{equation}
The equality in equation (\ref{bin2}) can be attained by the singlet state.
Historically, Bell-CHSH inequalities were the first tool for the recognition of entanglement; however, it is well-known for some time that the violation of a Bell-CHSH inequality is only a sufficient condition for entanglement and not a necessary one, and that there are in fact many entangled states that satisfy them \cite{werner}.
Bell-CHSH inequalities were generalized to N qubits, whose violations provide a criterion to distinguish the totally separable states from the entangled states \cite{gbin1,gbin2, gbin3}.

\item {\it Schmidt decomposition criterion}

For pure states there is a very simple necessary and sufficient criterion for separability, the Schmidt decomposition.  It allows one to write any pure state of a bipartite system as a linear combination of bi-orthogonal product states or, equivalently, of a non-superflous set of product states built from local bases \cite{sch, eker, peres, peres1}. Let us first recall the following.

{\it Theorem: Consider quantum systems $A$ and $B$ with dimensions $d_A,\: d_B$ respectively, and let $d = \;\textrm{min}\;(d_A, d_B)$. Any pure bipartite state $|\psi\rangle_{AB}$ has a Schmidt decomposition
\begin{equation}\label{sch1}
|\psi\rangle_{AB} = \sum_{i =1}^d\lambda_i|u_i\rangle_A|\nu_i\rangle_B,
\end{equation}
where $\lambda_i \ge 0$ and $\{ |u_i\rangle_A\}_i,\:\{ |u_i\rangle_B\}_i $ are 
orthogonal sets. The coefficients $\lambda_i$ are called the Schmidt coefficients and $|u_i\rangle_A,\:|\nu_i\rangle_B$, the Schmidt vectors.}

The Schmidt coefficients are uniquely defined and the number of non-zero Schmidt coefficients is called the Schmidt rank, defined as follows. For any bipartite pure state with Schmidt decomposition of the form (\ref{sch1}), the Schmidt rank is defined as the number of non-zero coefficients 
$\lambda_i$. A corollary of the Schmidt decomposition theorem is the following.

{\it A pure state in a composite system is a product state if and only if the Schmidt rank is 1, and is an entangled if and only if the Schmidt rank is greater than 1.}

{\it Example 1:} let us consider in two qubits system, where $\mathcal{H}_A$ and $\mathcal{H}_B$ are span by $\{ |0\rangle,\: |1\rangle\}$, the state 
\begin{equation}
|\psi\rangle_{AB} = \frac{1}{\sqrt 2}|0\rangle_A|1\rangle_B + \frac{1}{\sqrt 2}|1\rangle_A|0\rangle_B.
\end{equation}
 It is easy to check that $(1/\sqrt 2)^2 + (1/\sqrt 2)^2 = 1$, and that the state has Schmidt rank $2$ and therefore entangled.
 
{\it Example 2:} We consider now 
\begin{equation}
|\psi\rangle_{AB} = \frac{1}{2}\left( 
|0\rangle_A|1\rangle_B + |1\rangle_A|1\rangle_B + |0\rangle_A|0\rangle_B 
+ |1\rangle_A|0\rangle_B \right).
\end{equation}
It is easy to check that 
\begin{equation}
|\psi\rangle_{AB} = \frac{1}{2}(|0\rangle_A + |1\rangle_A)\otimes 
(|1\rangle_B + |0\rangle_B) = |+\rangle_A|+\rangle_B.
\end{equation}
Here $|+\rangle_A = \frac{1}{\sqrt 2}(|0\rangle_A + |1\rangle_A) $ and 
$|+\rangle_B = \frac{1}{\sqrt 2}(|0\rangle_B + |1\rangle_B) $.
In this example, the Schmidt rank is $1$, then $|\psi\rangle_{AB}$ is separable.

\item {\it Entropy of entanglement criterion}

A good way of characterizing the degree of entanglement of pure states is to measure the entanglement entropy, $S_A$,  that is for a state $|\psi\rangle_{AB}$, as in equation (\ref{sch1}), 
\begin{equation}\label{eent}
S_A = -\sum_{j=1}^d \lambda_j^2\;\textrm{log} \lambda_j^2 = -\textrm{Tr}_A\;\rho_A\;\textrm{log} \rho_A.
\end{equation}
The entanglement entropy, $S_B$, is determined as in the equation (\ref{eent}).

{\it The entanglement entropy is nonzero for entangled states and maximal when all the Schmidt coefficients $\lambda_j$ are equal. It is basis independent.}

We consider now a $2$ qubits system. Suppose the system is in the pure state 
\begin{equation}
|\psi\rangle = \frac{1}{\sqrt 2}(\;|00\rangle + |11\rangle\;), 
\end{equation}
so the density operator is $\rho_{AB} = |\psi\rangle\langle \psi |$. The reduced density matrix of system $A$ is 
\begin{eqnarray}\nonumber
\rho_A &=& \textrm{Tr}_B\rho_{AB};\\\nonumber
 &=& \frac{1}{2}{}_B\langle 0| (\; |00\rangle + |11\rangle\;)(\;\langle 11| + \langle 00|\;)|0\rangle_B\\\nonumber
 &{}& + \frac{1}{2}{}_B\langle 1| (\; |00\rangle + |11\rangle\;)(\;\langle 11| + \langle 00|\;)|1 \rangle_B;\\\nonumber
 & = & \frac{1}{2}(|0\rangle_A{}_A\langle 0| + |1\rangle_A{}_A\langle 1|).\\\nonumber
 & = & \frac{1}{2}\mathbb{I}_{2}.
\end{eqnarray}
The entanglement entropy of subsystem $A$ can be calculate using the formula in the equation (\ref{eent}) as follows
\begin{eqnarray}\nonumber
S_A &=& -\textrm{tr}\rho_A\;\textrm{log}\;\rho_A; \\\nonumber
&=&  -2\times \frac{1}{2}\;\textrm{log}\;\frac{1}{2};\\\nonumber
& =& \textrm{log}\;{2} \neq 0.
\end{eqnarray}

\item {\it The Positive Partial Transpose (PPT) criterion}

This criterion is also called the Peres-Horodecki criterion.
It was first proposed as a necessary condition for every separable state by A. Peres in $1996$ \cite{peres2}. He noticed that the separable states remain positive if subjected to partial transposition and then conjectured that this is also a sufficient condition. Later, M. Horodecki and al. studied the criterion in details and discovered that Peres's conjecture is valid for separable states of $2\times 2$ and $2 \times 3$ dimensions \cite{horo, phoro}. Thus for two qubits, the PPT criterion can be used to confirm exactly whether a state is entangled or not. The criterion can be formulated as follows.

{\it  Let $\rho_{AB} = \rho_{i\mu, j\nu}$ the density matrix of two qubits A and B. The entries of the density matrix that is partially transposed with respect to A are given by 
\begin{equation}
\rho^{T_A} = \rho_{j\mu,i\nu},
\end{equation}
where $\rho_A = \rho_{ij}$ and $\rho_B = \rho_{\mu\nu}$.} 

The theorem describing the criterion is as follows.

{\it Theorem: The partial transpose of a separable state $\rho_{AB}$ with respect to any subsystem is positive.}

For $2 \times 2$ and $2 \times 3$ systems $\rho_{AB}$ is separable if and only if $\rho^{T_A} $ is positive. The demonstration of the theorem can be found in \cite{peres1, horo, phoro}.

As an illustration, we consider a Werner state
\begin{equation}
\rho = p \;|\phi\rangle\langle \phi| + \frac{1-p}{4}\mathbb{I}_{4\times 4},
\end{equation}
where $0 < p < 1$ and $|\phi \rangle = \frac{1}{\sqrt{2}}(\;|0 1\rangle + |1 0\rangle\;)$. For $p = 0$ or $p= 1$, $\rho$ becomes a pure state. Explicitly we have 
\begin{eqnarray}\nonumber
\rho &=& \frac{p}{2}\left(\;  
(|0\rangle\langle 0| \otimes |1\rangle\langle 1|) +
(|0\rangle\langle 1| \otimes |1\rangle\langle 0|) +
(|1\rangle\langle 0| \otimes |0\rangle\langle 1| )+
(|1\rangle\langle 1| \otimes |0\rangle\langle 0|) \;\right)\\
&+& \frac{1-p}{4}(|0\rangle\langle 0| \otimes |0\rangle\langle 0 | 
+ |1\rangle\langle 1| \otimes |1\rangle\langle 1| + 
|0\rangle\langle 0| \otimes |1\rangle\langle 1| +|1\rangle\langle 1| \otimes |0\rangle\langle 0|).
\end{eqnarray}
For the calculations, it is better to write $\rho$ in matrix form as follows
\begin{equation}
|0\rangle = \left(\begin{array}{c}
1\\0
\end{array}\right);\quad \langle 0| = (1\:\:0); \quad |1\rangle = \left(\begin{array}{c}
0\\1
\end{array}\right);\quad \langle 1 | = (0\;\; 1).
\end{equation}
It is easy to find
\begin{equation}
\rho = \frac{1}{4}\left( 
\begin{array}{cccc}
1-p & 0 & 0 & 0\\
0 & p+1 & 2p & 0 \\
0 & 2p & p+1 & 0\\
0 & 0 & 0 & 1-p
\end{array}
\right).
\end{equation}
This density matrix has to be positive definite, all of its eigenvalues should be positive. The eigenvalues of $\rho$ are given by $\frac{1-p}{4},\;\frac{1-p}{4},\;  \frac{1-p}{4},\; \frac{1+ 3p}{4}$. For $0 <p <1$, they are all positive.

 Let's consider A the first subsystem and B the second subsystem. Let's apply the transpose to A only, that is called partial transpose to A, 
 \begin{eqnarray}\nonumber
 \rho^{T_A} &=& \frac{p}{2}\left( \; |0\rangle\langle 0 | \otimes |1\rangle\langle 1| )  
 + (|1\rangle\langle 0| \otimes |1\rangle\langle 0 |) + (|0\rangle\langle 1| \otimes |0\rangle\langle 1|) + (|1\rangle\langle 1|\otimes |0\rangle\langle 0|\right)\\
 &+& \frac{1-p}{4}\;\mathbb{I}_{4}.
 \end{eqnarray}
 In the matrix form we have 
 \begin{equation}
 \rho^{T_A} = \frac{1}{4}\left(
\begin{array}{cccc}
1-p & 0 & 0 & 2p \\
0 & p+1 & 0 & 0 \\
0 & 0 & p+1 & 0\\
2p & 0 & 0 & 1-p
\end{array} 
\right).
\end{equation}
According to Peres-Horodecki criterion, $\rho$ is separable if and only if both
 $\rho$ and $\rho^{T_A}$ are positive semi-definite, that means their eigenvalues are all positive. In the case there exist at least one negative eigenvalue then $\rho$ is entangled. The matrix $\rho^{T_A}$ has 3 eigenvalues that are equal to $\frac{p+1}{4} $ and another lowest eigenvalue $\frac{1-3p}{4}$, $0 < p< 1$. In order to have both eigenvalues of $\rho^{T_A}$ positive, p should be ranged between $0$ and $1/3$, that is ($0 < p < 1/3$ ). For instance for $p = 2/3$, we have the eigenvalues $\frac{5}{12};\; \frac{5}{12};\; \frac{5}{12};\; -\frac{1}{4}$. The fact that one eigenvalue is negative violate the theorem and therefore for $p = 2/3$ the state $\rho$ is entangled.
The PPT criterion is so strong that it characterizes entanglement for $2\times 2$
and $2\times 3$ systems, which means necessary and sufficient for qubit-qubit and qubit-qutrit systems only.

\item {\it The reduction criterion}

Applying the positivity criterion to the positive map $\Lambda (\sigma) = \mathbb{I}\; \textrm{Tr}\sigma - \sigma$ (with respect to the subsystems A and B), the following criterion was been provided \cite{horo, cerf}.

{\it Proposition: For any separable state $\rho_{AB}$, $\rho_A\otimes \mathbb{I}_B - \rho_{AB} \ge 0$, $\mathbb{I}_A \otimes \rho_B -\rho_{AB} \ge 0$.}
 We consider a density operator $\rho_{AB}$ of a bipartite system AB. The reduced density operator for system A is defined by $\rho_A = \textrm{Tr}_B(\rho_{AB})$ where $\textrm{Tr}_B$ is a map of operators known as the partial trace over the system B. The partial trace is defined as 
 \begin{equation}
 \textrm{Tr}_B(|a_1\rangle\langle a_2|\otimes |b_1\rangle\langle b_2|) = 
 |a_1\rangle\langle a_2|\; \textrm{Tr}_B(|b_1\rangle\langle b_2|)) = |a_1\rangle\langle a_2|\langle b_2|b_1\rangle.
 \end{equation}
 More generally, assuming $\rho_{AB} = \sigma\otimes \tau$, where $\sigma$ is the density operator for system $A$ and $\tau$ the density operator for system $B$, 
 \begin{equation}
 \textrm{Tr}_B(\rho_{AB}) = \textrm{Tr}_B(\sigma\otimes \tau) = \sigma \textrm{Tr}\tau = \sigma.
 \end{equation}
 In the same manner, $\textrm{Tr}_A(\rho_{AB}) = \tau$. The reduction criterion of separability is stated as follows.

 {\it Theorem: If $\rho_{AB}$ is separable then 
\begin{equation}
\rho_A\otimes \mathbb{I}_B -\rho_{AB} \ge 0,\: \textrm{and}\:\: \mathbb{I}_A\otimes \rho_B - \rho_{AB} \ge 0\;,
\end{equation}
where $\rho_A$ is the reduced matrix for system $A$ and $\rho_B$ the reduced density of system $B$.} 

Some simple examples are given as follows:

Example 1: We consider the state $|\psi\rangle_{AB} = |++\rangle_{AB} = |+\rangle_A|+\rangle_B$, the density matrix is then $\rho_{AB} = |++\rangle_{AB}\langle ++|_{AB} = (|+\rangle_A\langle +|_A) \otimes (|+\rangle_B\langle +|_B) $. In the basis 
$\{ |+\rangle,\;|-\rangle \}$. Let's determine now the reduced matrices $\rho_A$ and $\rho_B$. 
\begin{equation}
\rho_A = \textrm{Tr}_B(\rho_{AB} )= |+\rangle_A\langle + |_A = \left( \begin{array}{cc}
1 & 0\\
0 & 0
\end{array}\right);\quad \rho_A \otimes \mathbb{I}_{2} = 
\left(\begin{array}{cccc}
1 & 0 & 0 & 0\\
0 & 1 & 0 & 0\\
0 & 0 & 0 & 0\\
0 & 0 & 0 &0
\end{array} \right).
\end{equation}
\begin{equation}
\rho_B = \textrm{Tr}_A(\rho_{AB} )= |+\rangle_B\langle + |_B = \left( \begin{array}{cc}
1 & 0\\
0 & 0
\end{array}\right);\quad \mathbb{I}_{2} \otimes \rho_B = 
\left(\begin{array}{cccc}
1 & 0 & 0 & 0\\
0 & 0 & 0 & 0\\
0 & 0 & 1 & 0\\
0 & 0 & 0 &0
\end{array} \right).
\end{equation}
Let's check the reduction criterion theorem. 
\begin{equation}
\rho_{AB} = \left(\begin{array}{cccc}
1 & 0 & 0 & 0\\
 0 & 0 & 0 & 0\\
 0 & 0 & 0 & 0\\
 0 & 0 & 0 & 0
\end{array}
\right),
\end{equation}
and 
\begin{equation}
 \rho_A\otimes \mathbb{I}_{ 2} -\rho_{AB} = 
\left(\begin{array}{cccc}
0 & 0 & 0 & 0\\
 0 &1 & 0 & 0\\
 0 & 0 & 0 & 0\\
 0 & 0 & 0 & 0
\end{array}
\right); \quad 
\mathbb{I}_{2}\otimes \rho_B -\rho_{AB} = 
\left(\begin{array}{cccc}
0 & 0 & 0 & 0\\
 0 &0 & 0 & 0\\
 0 & 0 & 1 & 0\\
 0 & 0 & 0 & 0
\end{array}
\right).
\end{equation}
$\rho_A\otimes \mathbb{I}_{ 2} -\rho_{AB} \ge 0$ 
and $\mathbb{I}_{ 2}\otimes \rho_B -\rho_{AB} \ge 0$ because it is easy to see that their eigenvalues are all greater or equal to zero, therefore the state is separable.

We consider now a second example that is the state $|\psi\rangle_{AB} = \frac{1}{\sqrt 2} (|0\rangle_A|0\rangle_B + |1\rangle_A|1 \rangle_B)$. The associated density matrix 
\begin{equation}
\rho_{AB} = |\psi\rangle_{AB}\langle \psi|_{AB} = \frac{1}{2}
\left(\begin{array}{cccc}
\begin{array}{cccc}
1 & 0 & 0 & 1\\
0 & 0 & 0 & 0\\
 0 & 0 & 0 & 0 \\
 1 & 0 & 0 & 1
\end{array}
\end{array} \right)
\end{equation}
\begin{equation}
\rho_A = \textrm{Tr}_B{\rho_{AB}} = \frac{1}{2}(\;|0\rangle_A\langle 0|_A + 
|1\rangle_A\langle 1 |_A\;) = \frac{1}{2}\left(\begin{array}{cc}
1 & 0 \\
0 & 1
\end{array} \right);
\end{equation}
we have 
\begin{equation}
\rho_A \otimes \mathbb{I}_{2} - \rho_{AB} = \frac{1}{2} 
\left(\begin{array}{cccc}
0 & 0 & 0 & 1\\
0 & -1 & 0 & 0\\
0 & 0 & -1  & 0 \\
1 & 0 & 0 & 0 
\end{array} \right).
\end{equation}
It is easy to check that $\lambda = -1/2$ is an eigenvalue of $\rho_A \otimes \mathbb{I}_{2} -\rho_{AB}$,  therefore the state $|\psi\rangle_{AB}$ 
is entangled since $\rho_A \otimes \mathbb{I}_{2} -\rho_{AB}$ is not positive.

A remark is that the reduction criterion is a necessary and sufficient separability condition only for the dimensions $2 \times 2$ and $2\times 3$ systems but it is just a necessary condition in higher dimensions.

An other remark is that there is another trick to recognize a pure entangled state related to reduced density is the following: if the reduced matrix $\rho_A$ for a pure state $\rho_{AB}$ is mixed then $\rho_{AB}$ is entangled. 
 
\item {\it Concurrence criterion}

The term concurrence may refer to different meanings according to Wikipedia. In quantum computing, concurrence refers to a measure of quantum entanglement. We also consider that measure as a way of characterizing entanglement. For a pure state $|\phi\rangle$ of a pair of qubits, the concurrence denoted $C(|\phi\rangle)$ is defined as 
\begin{equation}
C(|\phi\rangle) = |\langle \phi| \tilde{\phi}\rangle|,
\end{equation}
where $|\tilde{\phi}\rangle = (\sigma_y\otimes \sigma_y)|\phi^*\rangle$, with $\sigma_y$ the Pauli operator $\left( \begin{array}{cc}
0 & -i\\
i & 0
\end{array}\right)$ and $|\phi^*\rangle$ the complex conjugate of $|\phi\rangle$. Explicitly 
\begin{equation}
|\tilde{\phi}\rangle = \left( 
\begin{array}{cccc}
0 & 0 & 0 & -1 \\
0 & 0 & 1 & 0 \\
0 & 1 & 0 & 0 \\
-1 & 0 & 0 & 0
\end{array}
\right)|\phi^* \rangle.
\end{equation}
The state $|\tilde{\phi}\rangle $ is called the ``spin flip" state of $|\phi\rangle$ \cite{woo1, woo2}. The spin flip operation, when applied to a pure product state, takes the state of each qubit to the orthogonal state. 

A more clear definition of concurrence establishing a connection with entanglement is given in \cite{woo3, woo2, zhao} as follows. A pure state $|\phi\rangle$ in a two-qubit system can be expressed in the standard basis as 
\begin{equation}
|\phi\rangle = \alpha|00\rangle + \beta |01\rangle + \gamma |10\rangle + \eta |11\rangle,
\end{equation} 
where $|\alpha|^2 + |\beta|^2 + |\gamma|^2 + |\eta|^2 = 1$. It can be shown that $|\phi\rangle$ is factorizable only under the case $\alpha\eta = \beta\gamma$. Therefore, the difference between $\alpha\eta$ and $\beta\gamma$ can be taken as a measurement of entanglement. In this way a definition of the concurrence is given by 
\begin{equation}
C(|\phi\rangle) = 2 |\alpha\eta -\beta\gamma|.
\end{equation}
From this definition, the concurrence is zero for separable state, and that is a criterion of separability.
For example for a two qubit state $|\phi\rangle = \frac{1}{2}(|00\rangle + |11\rangle)$, the concurrence is given by 
$ C(|\phi\rangle) = 2 |\;\frac{1}{\sqrt 2}\times \frac{1}{\sqrt{2}} - 0\times 0 \:| = 1$.

The concurrence of a mixed state $\rho$ of two qubits can be defined as the average concurrence of an ensemble of pure states representing $\rho$, minimized over all decomposition of $\rho$:
\begin{equation}
C(\rho) ={ \textrm{inf}}_{\{p_j,\; |\phi_j\rangle\}}\lbrace\sum_j\;p_jC(\phi_j)\rbrace,
\end{equation}
where $\rho = \sum_j p_j|\phi_j\rangle\langle \phi_j |$ and the states $|\phi_j\rangle$ are distinct normalized pure states of the bipartite system, not necessarily orthogonal. As a consequence, a state $\rho$ is separable if and only if $C(\rho) = 0$.

An explicit formula for concurrence is given by  \cite{woo3} as 
\begin{equation}
C(\rho) = \textrm{max}\;\{0, \lambda_1 -\lambda_2 -\lambda_3 -\lambda_4\},
\end{equation}
where $\lambda_i (i = 1,2,3,4)$ is the non-negative eigenvalue of the Hermitian matrix $R = \sqrt{\sqrt{\rho}\tilde{\rho}\sqrt{\rho}}$, in decreasing order.
Here $\tilde{\rho} = (\sigma_y\otimes\sigma_y)\rho^*(\sigma_y\otimes \sigma_y)$, where $\rho^*$ is the complex conjugate of $\rho$, and as a mixed state $\rho = \sum_jp_j|\phi_j\rangle\langle \phi_j|$. 

\item {\it The majoration criterion}

We start by defining what we mean by majoration. Consider two d-dimensional real vectors 
\begin{equation}
 x = (x_1,x_2, \ldots, x_d)\quad \textrm{and} \quad y = (y_1,y_2,\ldots y_d).
\end{equation}
It is often assumed in addition that $x$ and $y$ are probability distributions, that is, the components are non-negative and sum to one as follows.
\begin{equation}
 x_i \ge 0, \quad 1 \le i \le d \quad \textrm{and }\quad \sum_{i = 1}^d x_i = 1;
\end{equation}
\begin{equation}
 y_i \ge 0, \quad 1 \le i \le d \quad \textrm{and }\quad \sum_{i = 1}^d y_i = 1.
\end{equation}
 Let's introduce now the notation $\downarrow$ to denote the components of a vector rearranged into decreasing order. So $x^{\downarrow} = 
 \left(x_1^{\downarrow}, x_2^{\downarrow}, \ldots, x_d^{\downarrow} \right)$ in the sense that $x_1^{\downarrow}\ge x_2^{\downarrow} \ge \ldots\ge  x_d^{\downarrow}$. The vector $x^{\downarrow}$ is majorized by the vector $y^{\downarrow}$, that is denoted $x^{\downarrow} \prec y^{\downarrow} $ when 
 \begin{equation}
 \sum_{j=1}^k x_j^{\downarrow} \le  \sum_{j=1}^k y_j^{\downarrow}, \quad 1 \le k \le d-1,
 \end{equation}
and the equality holds for $k = d$, with $d$ being the dimension of the vectors.
More details about the majoration can be found in the references \cite{marshall, nielsen}. 
We consider now the density matrix $\rho_{AB}$ of a pair of qubits. Let $\rho_A$ be the reduced density matrix of the system of the qubit A and $\rho_B$ the reduced density matrix of qubit $B$. If $\lambda_{\rho_{AB}}$ consist of the eigenvalues of $\rho_{AB}$ and $\lambda_{\rho_A}$, $\lambda_{\rho_B}$  the eigenvalues of $\rho_A$, $\rho_B$ respectively, 

{\it the majoration criterion of separability says that if the state $\rho_{AB}$ is separable then 
\begin{equation}\label{maj}
\lambda_{\rho_{AB}}^\downarrow\prec \lambda_{\rho_A}^\downarrow, \quad 
\textrm{and} \quad \lambda_{\rho_{AB}}^\downarrow\prec \lambda_{\rho_B}^\downarrow.
\end{equation}}
Zeros are appended to the vectors $ \lambda_{\rho_A}^\downarrow$ and $ \lambda_{\rho_B}^\downarrow$ in the equation(\ref{maj}) in order to make their dimensions equal to the one of $\lambda_{\rho_{AB}}$.
For a separable state the ordered vector of eigenvalues of the whole density operator is majorized by the ones of the reduced density matrices \cite{nielsen}.
An important remark is that the spectra of a density matrix and its reduced density matrices do not allow to distinguish separable and entangled states. The majoration criterion is only a necessary and not a sufficient condition of separability.  For the system of two qubits the operational separability criteria follows this logical ordering:

{\it $\rho_{AB}$ separable $\Leftrightarrow$ $\rho_{AB}$ satisfies the reduction criterion $\Rightarrow$ $\rho_{AB}$ satisfies the majoration criterion.}

As illustration we consider the Werner state
\begin{equation}
\rho_{AB} = p|\phi\rangle\langle \phi| + \frac{1-p}{4}\mathbb{I}_{4},
\end{equation}
where $0 \le p \le 1$ and $|\phi \rangle = \frac{1}{\sqrt{2}}(|01\rangle + |10\rangle)$. This state has already been studied and its matrix matrix  form is 
\begin{equation}
\rho_{AB} = \frac{1}{4}\left( 
\begin{array}{cccc}
 1-p & 0 & 0 & 0 \\
 0 & p+1 & 2p & 0 \\
 0 & 2p & p+1 & 0 \\
 0 & 0 & 0 & 1-p
\end{array}
\right).
\end{equation}
Its eigenvalues are 
\begin{equation}
\lambda_{\rho_{AB}}^\downarrow = \left\lbrace
\frac{3p +1}{4},\; \frac{1-p}{4},\; \frac{1-p}{4},\;
\frac{1-p}{4}\right\rbrace, \: 0\le p \le 1.
\end{equation}
The reduced density matrix $\rho_A$ and the reduced density matrix $\rho_B$
are respectively given in matrix form as 
\begin{equation}
\rho_A  = \textrm{Tr}_B(\rho_{AB}) = \frac{1}{2}\left(
\begin{array}{cc}
1 & 0 \\
0 & 1
\end{array}
\right) = \rho_B.
\end{equation}
The eigenvalues of $\rho_A $ and $\rho_B$ respectively appended by zeros 
are given respectively by 
\begin{equation}
\lambda_{\rho_A}^\downarrow  =  \left\lbrace \frac{1}{2},\; \frac{1}{2},\; 0,\; 0\right\rbrace; \quad 
\lambda_{\rho_B}^\downarrow  = \left\lbrace \frac{1}{2},\; \frac{1}{2},\; 0,\; 0\right\rbrace, \quad 0  \le p \le 1.
\end{equation}
All conditions are fulfilled to check the separability condition using the majoration criterion. We know that $0 \le p \le 1/3 $, is separable according to the PPT criterion and this is fulfilled by 
\begin{equation}
\sum_{j = 1}^k \lambda_{\rho_{AB},j}^\downarrow \le \sum_{j = 1}^k \lambda_{\rho_{A},j}^\downarrow, \quad \textrm{and} \quad 
\sum_{j = 1}^k \lambda_{\rho_{AB},j}^\downarrow \le \sum_{j = 1}^k \lambda_{\rho_{B},j}^\downarrow
\end{equation}
so
\begin{equation}
\lambda_{\rho_{AB}}^\downarrow \prec \lambda_{\rho_{A}}^\downarrow\quad \textrm{and}\quad
\lambda_{\rho_{AB}}^\downarrow \prec \lambda_{\rho_{B}}^\downarrow
\end{equation}
One can realize that for $p > 1/3$ the majoration condition is not satisfied.

\item {\it The computable cross norm or realignment (CCNR) criterion}

For finite-dimensional systems, a criterion for the separability  is the so-called computable cross norm or realignment [CCNR] criterion proposed by 
Rudoph \cite{rudolph1, rudolph2}, Chen and Wu \cite{chenwu}. 

The CCNR criteria has been found in two different forms, that are, the computable cross norm criterion by Rudolph and the the realignment criterion by Chen and Wu.
The separability criterion is given either by defining a new norm (a cross norm) or by realigning the density operator and them taking the usual trace norm of the realigned matrix.  

\begin{enumerate}
\item The computable cross norm (CCN) criterion is an analytical and computable separability criterion for bipartite quantum states developped by O. Rudolph \cite{rudolph1}, known to systematically detect bound entanglement and complements in certain aspects the well-known Peres positive partial transpose (PPT) criterion. It can be formulated in different equivalent ways. A very useful and instructive way is the following procedure \cite{rudolph3}. Consider a quantum state $\rho_{AB}$ defined on  a tensor product Hilbert space 
$\mathcal{H}_A \otimes \mathcal{H}_B$. We denote the canonical real basis  $(|i\rangle)_{i}$ and expand $\rho_{AB}$ in terms of the operators $E_{ij} = |i\rangle\langle j|$, we write 
\begin{equation}
\rho_{AB} = \sum_{ijkl}\rho_{ijkl} E_{ij}\otimes E_{kl}.
\end{equation}
Next, we define an operator $\mathcal{U}(\rho_{AB})$ that acts on $T(\mathcal{H}_A \otimes \mathcal{H}_B)$ by 
\begin{equation}
\mathcal{U}(\rho_{AB}) = \sum_{ijkl}\rho_{ijkl}|E_{ij}\rangle\langle E_{kl}|,
\end{equation}
where $T(\mathcal{H}_A\otimes \mathcal{H}_B)$ denotes the trace class operators on $\mathcal{H}_A\otimes \mathcal{H}_B$. Here, the ope\-ra\-tors $|E_{ij}\rangle$ denotes the ket vector with respect to Hilbert Schmidt inner product $\langle A, B\rangle \equiv \textrm{Tr} (A^\dagger B)$ in $T(\mathcal{H}_A\otimes \mathcal{H}_B)$.
We also write $||A||_2 \equiv \langle A, \; A\rangle^{1/2}$. The norm $||A||_2$
is often called the Hilbert-Schmidt norm or the Frobenius norm of A and is equal to the sum of the squares of the singular values of $A$. The sum of the absolute values of the singular values of $A$ is called the trace class norm, or simply trace norm, and is denoted by $||A||_1$.  The CCN criterion is then formulated as follows.

{\it The CCN criterion asserts that if $\rho$ is separable, then the trace norm of $\mathcal{U}(\rho) $ is less than or equal to one. Whenever a quantum state $\rho$ satisfies $||\mathcal{U}(\rho)||_1 > 1$, this signals that $\rho$ is entangled.}

Let's consider the following two qubit example. We consider two Hilbert spaces $\mathcal{H}_A$ and $\mathcal{H}_B$ span by $\{ |0\rangle, |1\rangle \}$, respectively. Next we consider the family of states on $\mathcal{H}_A \otimes \mathcal{H}_B$
\begin{equation}\label{ccn}
\rho_p = p|00\rangle\langle 00 | + (1-p)|\phi\rangle\langle \phi|,
\end{equation}
where $0\le p \le 1$ and $|\phi\rangle = \frac{1}{\sqrt 2}(|01\rangle + |10\rangle)$. The matrix operator in equation (\ref{ccn}) can be expanded as follows
\begin{eqnarray}\nonumber
\rho_p &=& p|0\rangle\langle 0| \otimes |0\rangle\langle 0| + \frac{1-p}{2}|0\rangle\langle 0| \otimes |1\rangle\langle 1| +
\frac{1-p}{2}|0\rangle\langle 1| \otimes |1\rangle\langle 0|\\
 & +& \frac{1-p}{2}|1\rangle\langle 0| \otimes |0\rangle\langle 1|
 +
\frac{1-p}{2}|1\rangle\langle 1| \otimes |0\rangle\langle 0|.
\end{eqnarray}
Defining now $E_{ij} = |i\rangle\langle j |$, we have 
\begin{equation}
\rho_p = p E_{00}\otimes E_{00} + \frac{1-p}{2}\left( 
E_{00}\otimes  E_{11} + E_{01}\otimes E_{10} + E_{10}\otimes E_{01} 
+ E_{11}\otimes E_{00}
\right).
\end{equation}
Next, we define an operator $\mathcal{U}(\rho_p)$ that acts on $T(\mathcal{H}_A\otimes \mathcal{H}_B)$ by 
\begin{equation}
\mathcal{U}(\rho_p) = p |E_{00}\rangle\langle E_{00}| + \frac{1-p}{2}\left(\;
|E_{00}\rangle\langle E_{11}| + |E_{11}\rangle\langle E_{00}| +
|E_{01}\rangle\langle E_{10}| + |E_{10}\rangle\langle E_{01}| 
\;\right),
\end{equation}
where $|E_{ij}\rangle$ denotes the ket vector with respect to Hilbert-Schmidt inner product. The trace class norm of $\mathcal{U}(\rho_p)$ can be computed as follows \cite{rudolph2}.
\begin{eqnarray}\nonumber
||\mathcal{U}(\rho_p||_1 &=& 1-p + \sqrt{\frac{p^2}{2} + \frac{(1-p)^2}{4} + 
\frac{p}{2}\sqrt{p^2 + (1-p)^2}}\\\nonumber
&{}& +\; \sqrt{\frac{p^2}{2} + \frac{(1-p)^2}{4} -
\frac{p}{2}\sqrt{p^2 + (1-p)^2}}\\
&\ge & 1-p + p\sqrt{1 + \frac{(1-p)^2}{2p^2}} \ge 1.
\end{eqnarray}
We have egality if and only if $p= 1$, so $\rho_p$ is separable if and only if 
$p = 1$.

The CCN criterion is in general not a sufficient criterion for separability in 
dimension $2 \times 2$. For two qubits states with maximally disordered subsystems the CCN criterion is necessary and sufficient. It is shown that for all pure states, for Bell diagonal states, for Werner states in dimension $ d= 2$, and for isotropic states in arbitrary dimensions, the CCN criterion is necessary and sufficient.

\item A matrix realignment criterion

Motivated by the Kronecker product approximation technique, a method to assess the inseparability of bipartite quantum systems, based on a realigned matrix contructed from the density matrix has been developed by Chen and Wu \cite{chenwu}. 
Let's define the realigned matrix as follows. 

We consider a bipartite system $A$, $B$ represented by two Hilbert space 
$\mathcal{H}_A$ and $\mathcal{H}_B$ of dimension $d_A$ and $d_B$ respectively. Let's consider now a density matrix $\rho_{AB}$ acting on $\mathcal{H}_A\otimes \mathcal{H}_B$, the realigned matrix $\mathcal{R}(\rho_{AB})$ is such that the matrix elements are
\begin{equation}
\langle a_i, b_l|\; \mathcal{R}(\rho_{AB})\; |b_k, a_j\rangle = \langle a_i, b_k|\; \rho_{AB}\;| a_j, b_l\rangle,
\end{equation}
where $\{ |a_i, b_k\rangle\}, \: i =  1, \ldots, d_A; k= 1, \ldots, d_B$, is a basis of $\mathcal{H}_A \otimes \mathcal{H}_B$. The realigned matrix is of dimension $d_A^2\times d_B^2$.

Consider a bipartite state $\rho_{AB}$ acting on $\mathcal{H}_A\otimes \mathcal{H}_B$ as 
\begin{equation}
\rho_{AB} = \sum_{i,j}\sum_{k,l}c_{ijkl}|a_i,b_k\rangle\langle a_j, b_l|,
\end{equation}
and its realigned matrix 
\begin{equation}
\mathcal{R}(\rho_{AB}) = \sum_{i,j}\sum_{k,l}c_{ijkl}|a_i,b_l\rangle\langle b_k, a_j|,
\end{equation}
 The realignment criterion based on the matrix $\mathcal{R}(\rho_{AB})$ states that if the state $\rho_{AB}$ is separable, then $||\; \mathcal{R}(\rho_{AB})\;||_1 \le 1$ must hold \cite{horo}.
\end{enumerate}
The CCN criterion and the realignment criterion are equivalent and known under computable cross norm or realignment criterion (CCNR).

{\it The CCNR criterion states that if $\rho_{AB}$ is a separable state on $\mathcal{H}_A \otimes \mathcal{H}_B$ with $\textrm{dim}(\mathcal{H}_A\otimes \mathcal{H}_B) < +\infty$, then the trace norm 
$|| \mathcal{R}(\rho_{AB})||_1$ of the realignment matrix $\mathcal{R}(\rho_{AB})$ of $\rho_{AB}$ is not greater than 1.}

\item {\it The correlation matrix (or de Vicente) criterion (2008)}

The correlation matrix (or de Vicente) criterion involves the Bloch representation of density operators . About Bloch operators (density matrices), one may read in \cite{block, hioe}. The criterion, developed in 2007 by de Vicente, is a necessary condition and can detect PPT entangled states \cite{vicente}. The correlation matrix criterion is formulated as follows.

In the case of $d_A \times d_B$ bipartite quantum systems $d_A \le d_B$, the Bloch representation, also known as Fano form \cite{fano} can be written as follows.
\begin{equation}
\rho_{AB} = \frac{1}{d_Ad_B}
\mathbb{I}_{d_A}\otimes \mathbb{I}_{d_B} + \frac{1}{2d_B}\vec{r}\cdot\vec\lambda\otimes \mathbb{I}_{d_B} + \frac{1}{2d_A}\mathbb{I}_{d_A}\otimes \vec{s}\cdot \vec\sigma + \sum_{i = 1}^{d_A^2-1}\sum_{j=1}^{d_B^2 -1}\tau_{ij}\lambda_i \otimes \sigma_j,
\end{equation}
where $\mathbb{I}_{d_A}$ and $\mathbb{I}_{d_B}$ are the identity matrices, $\vec r$ and $\vec \sigma$ have the components $r_i = \textrm{Tr}[\rho_{AB}(\lambda_i\otimes \mathbb{I}_{d_B})],\: i = 1,2,\ldots d_A^2-1$ and $s_j = \textrm{Tr}[\rho_{AB}(\mathbb{I}_{d_A}\otimes \sigma_j)],\: j= 1,2,\ldots d_B^2 -1 $, and the correlation matrix $\tau_{ij} = \textrm{Tr}[\rho_{AB}(\lambda_i\otimes \sigma_j)]$. The vector $\vec\lambda$ is defined to be $\lambda \equiv (\lambda_1, \lambda_2 \ldots \lambda_{d_A^2-1})^T$, with $\lambda_i$, being the generators of $SU(d_A)$. The vector $\vec\sigma $ is defined similarly with $\sigma_j$, being the generators of $SU(d_B)$. Here $\vec{r}$ and $\vec{s}$ are called the Bloch vectors of the density matrices.

{\it If the state is separable, then the inequality \cite{vicente}, 
\begin{equation}\label{vicentec}
||\tau||_1 \le \sqrt{\frac{4 (d_A -1)(d_B -1)}{d_Ad_B}}.
\end{equation}
must hold, with $\tau$ the matrix corresponding to the matrix elements 
$\tau_{ij}$. }

The matrix $\tau$ accounts for the possible correlations between the subsystems and is therefore called the correlation matrix. Note that if $\tau = 0$, then the state $\rho$ is separable, but the converse is not true. Physically, this necessary condition means that there is an upper bound to the amount of correlation contained in a separable state, that means that the correlations in separable states cannot be `` too large". For the case of two qubits system with maximally mixed reduced density matrix, the criterion is necessary and suffiscient 
in reference to example 2 in \cite{vicente}.

We have $d_A = d_B = 2$, so 
\begin{equation}
\rho_{AB} = \frac{1}{4}\left(
\mathbb{I}_{2}\otimes \mathbb{I}_{2} + \sum_{i = 1}^{3}r_i\lambda_i\otimes \mathbb{I}_{2} + 
\sum_{j = 1}^{3}s_j\mathbb{I}_{2}\otimes \sigma_j + \sum_{i = 1}^{3}\sum_{j=1}^{3}\tau_{ij}\lambda_i \otimes \sigma_j
\right),
\end{equation}
where $\{\lambda_i\}_{i=1}^3$, $\{\sigma_i\}_{i=1}^3$ denotes the generators of $SU(2)$, and $r_i = \textrm{Tr}(\lambda_i\otimes \mathbb{I}_2\;\rho_{AB}),\;s_i = \textrm{Tr}(\mathbb{I}_2\otimes \sigma_i\;\rho_{AB}),\; i=1,2,3$. The matrix $\tau$ is given by $\tau_{ij} = \textrm{Tr}(\lambda_i\otimes \sigma_j)\rho_{AB},\; i, j = 1,2,3$. The three generators of $SU(2)$ are Pauli matrices 
\begin{equation}
\lambda_1 = \left( \begin{array}{cc}
0 & 1\\
1 & 0
\end{array}\right) = \sigma_1; \quad \lambda_2 = \left( \begin{array}{cc}
0 & -i\\
i & 0
\end{array}\right) = \sigma_2;\quad \lambda_3 = \left( \begin{array}{cc}
1 & 0\\
0 & -1
\end{array}\right) = \sigma_3.
\end{equation}

The separability criterion (\ref{vicentec}) for the case of two qubits becomes 
\begin{equation}
||\tau||_1 \le 1.
\end{equation}
\item {\it Enhanced entanglement criterion via SIC POVMs}

This criterion is based on a special measurement called the symmetric informationally complete positive operator valued measures (SIC POVMs) proposed in 2018 by Shang, Asadian, Zhu and G\"uhne \cite{shang}. Let's recall first the definition of positive-operator valued measure (POVM) as follows.

{\it A set of operator $\{ O_i,\; i\in M \}$ is called a POVM if each operator $O_i$ satisfies the following properties: 1) Hermitian: $O_i = O_i^\dagger$; 2) $O_i$ is positive semi-definite: $O_i\ge 0$; 3) $\{O_i,\; i\in M\}$ forms a resolution of the identity on M: $\sum_{i\in M}O_i = \mathbb{I}_M$. The probability of obtaining outcome i for a given state specified by density matrix $\rho$ is given by 
\begin{equation}
p_i = \textrm{Tr}(O_i\rho).
\end{equation}}
A POVM in a $d$-dimensional Hilbert space is called informationally complete if the probabilities $p_i$ uniquely determine the density operator, which means that it must contain at least $d^2$ elements in order to span the Hilbert Schmidt space $\mathcal{H}\mathcal{S}(\mathcal{H})$, with $d$ the dimension of $\mathcal{H}$ \cite{icpovm}.

The POVM is said to be symmetric informationally complete (SIC) known as SIC POVMs if it is composed of $d^2$ elements 
\begin{equation}
\Pi_k = \frac{1}{d}|\psi_k\rangle\langle \psi_k|,
\end{equation}
which are subnormalized rank $1$ projectors onto pure states 
with equal pairwise fidelity, that is
\begin{equation}
|\langle \psi_k|\psi_l\rangle|^2 = \frac{d\;\delta_{kl} + 1}{d + 1}, \quad 
k, l = 1, \ldots d^2,
\end{equation}
and satisfying the completeness condition $\sum_{k=1}^{d^2}\Pi_k = \mathbb{I}$.
It is still a conjecture that SIC POVMs exist in all finite dimensions.

Given a SIC POVM $\mathcal{M}_s = \{\Pi_k\}_{k = 1}^{d^2}$ and a quantum state $\rho$, the probability of obtaining outcome $k$ is given by the Born rule, 
$p_k = \langle \pi_k\rangle = \textrm{Tr}(\rho\Pi_k)$.

Consider now a bipartite state $\rho$ acting on the Hilbert space $\mathcal{H}_{AB} = \mathcal{H}_A \otimes \mathcal{H}_B$ with dimension $d_{AB} = d_A \times d_B$, and let $\{ E_k^A\}_{k=1}^{d_A^2}$ and $\{E_k^B\}_{k= 1}^{d_B^2}$ be normalized SIC POVMs for the two respective subsystems: the linear correlations between the two SIC POWMs $E^A$ and $E^B$ are 
\begin{equation}
[P_s]_{kl} = \textrm{Tr}( E_k^A \otimes E_l^B\; \rho)
\end{equation}
from which the entanglement criterion based on the SIC POVMs, called ESIC criterion is as follows.

{\it If a state $\rho $ is separable, then  $||P_s||_1 \le 1$ has to hold, otherwise, it is entangled.}

While the existence of SIC-POVMs for all dimensions is yet to be proven, there are analytical and/or numerical solutions for dimensions up to 151. 
\end{enumerate}

\section{Non operational criteria}\label{sec3}

Non operational separability criteria do not provide us with a simple procedure to check the separability properties of a given state. So they are not easy to use. However, they are necessary and sufficient criteria for any bipartite system.
\begin{enumerate}

\item {\it Positive maps criterion }

A standard approach to decide on the separability of a given mixed state relies on positive maps. Quantum entanglement theory is linked with
the theory of positive maps \cite{pm1, pm2, pm3, pm4, pm5}. A map $\Lambda: \mathcal{B}(\mathcal{H})\rightarrow \mathcal{B}(\mathcal{H}))$ is called positive if it maps positive operators on positive ones, means 
\begin{equation}
\Lambda(\rho) \ge 0, \quad \textrm{for all}\quad \rho \ge 0\:,
\end{equation}
where the positivity of an operator $\rho$ is to state that $\rho$ is positive semi-definite  that means $\rho$ has only non-negative eigenvalue. Here $\mathcal{B}(\mathcal{H})$ denotes the set of bounded operators on $\mathcal{H}$. A crucial property of positive maps is that a trivial extension $\Lambda \otimes \mathbb{I}$ is not necessarily positive and this fact can be used to conclude on the separability of a mixed state $\rho$ acting on $\mathcal{H}_A\otimes \mathcal{H}_B $. 
The positivity of $\Lambda$ means that $\Lambda (X)\ge 0$ for any $ X\ge 0$. Recall that $X$ is positive (which is denoted by $X \ge 0$) if and only if
$\langle \psi | X| \psi \rangle \ge 0$ for any vector $|\psi\rangle$. This is equivalent to the requirement that $X$ is Hermitian operator with nonnegative spectrum \cite{horo}. The positive maps criterion of separability is as follows.

{\it A state $\rho_{AB}$ defined on $\mathcal{H}_A \otimes \mathcal{H}_B$ is separable if and only if for all possible linear maps
\begin{equation}
\Lambda : \mathcal{B}(\mathcal{H}_B)\rightarrow \mathcal{B}(\mathcal{H}_A)
\end{equation}
one has $(\mathbb{I}\otimes \Lambda )\rho_{AB}\ge 0$.}

While the separability of pure states can easily be checked, it turns out to be much difficult to decide whether a given mixed state is entangled or separable.
This statement does not allow to derive a sufficient separability criterion for the very general case, since the classification of positive maps is still an unsolved problem.

\item {\it The entanglement witnesses}

There is natural characterization of separable states in terms of mean
values of Hermitian operators \cite{horo, terhal1}.
The scalar separability criteria are represented as some bounds on values of scalar functions of the density matrix $\rho$ \cite{terhal1, kraus}. The term entanglement witness was first used by Terhal in 2000 \cite{terhal1}, and refers to operators that can detect entangled states. The definition of entanglement witness is as follows.

A Hermitian operator $W$ acting on a bipartite system $\mathcal{H}_{AB}$ is called an entanglement witness if it satisfies the following properties:
\begin{equation}
\langle \phi\otimes \psi|\;  W \;| \phi\otimes \psi\rangle \ge 0, \quad \forall\quad |\phi\rangle \otimes |\psi\rangle \in \mathcal{H}_{AB};
\end{equation}
\begin{equation}
\exists\:|\chi\rangle \in \mathcal{H}_{AB},\; \langle \chi\ |\: W \:|\chi \rangle < 0.
\end{equation}
The criterion based on entanglement witness is as follows.

{\it Proposition: The state $\rho$ is separable if and only if $\textrm{Tr}(\rho W)\ge 0$ for all Hermitian operators W called entanglement witnesses (EW) such that 
\begin{itemize}
\item $\textrm{Tr}\;(\sigma W)\ge 0$ for all separable $\sigma$;
\item there exists some entangled $\rho$ such that $\textrm{Tr}\;(W\rho)< 0$.
\end{itemize}}
Any fixed entanglement witness $W$ provides necessary condition for separability $\textrm{Tr}(W\rho) \ge 0$. The first explicit example of operators satisfying the properties in the proposition was provided in \cite{werner} as follows.

This was ``flip operator" $V$ defined for $d \otimes d$ systems as 
$V |\phi\rangle |\psi\rangle = |\psi\rangle|\phi \rangle$ for all $|\phi\rangle,\: \psi\rangle \in C^d$. It reveals entanglement for $|\psi_-\rangle$ as 
$\textrm{Tr} (V|\psi_- \rangle\langle \psi_-|) = -1 < 0$, with $|\psi_-\rangle = \frac{1}{\sqrt 2}(|01\rangle - |10\rangle)$.

The separability criterion based on entanglement witnesses is a necessary and sufficient criteria but not practical. Some widely encountered entanglement witnesses and their properties are given in \cite{terhal1, guhne, ganguly, dagmar}. The geometric entanglement witnesses are discussed in \cite{kammer}, and are of great use.
 
\item {\it Local uncertainty relations (LURs) criterion}

The local uncertainties relations (LURs) by Hofmann and Takeuchi \cite{lur1} are based on the reformulation of the uncertainty principle, adapting it to arbitrary properties of $N$-level systems and providing unconditional limitations for the predictabilities of measurement outcomes for any selection of non-commuting physical properties. Local uncertainty limits valid for all non-entangled states can be then derived. Since no separable quantum state can overcome these limits, any violation of such uncertainty relations is an unambiguous proof of entanglement. The local uncertainty relations (LURs) criterion is formulated as follows.

{\it We consider a bipartite system AB represented by the Hilbert space $\mathcal{H}_A$ and $\mathcal{H}_B$. Given some non-commuting observables 
$A_k$ on $\mathcal{H}_A$ and $B_k$ on $\mathcal{H}_B$, one may compute strictly positive numbers $C_A$  and $C_B$ such that 
\begin{equation}
\sum_{k=1}^n\Delta^2(A_k)\ge C_A\;, \quad \sum_{k=1}^n\Delta^2(B_k)\ge C_B
\end{equation}
holds for all states for system A, respectively system B.}
Here, 
\begin{equation}
\Delta^2(A) =\; <A^2> - <A>^2
\end{equation}
 denotes the variance of an observable A. It can then be proved that 
for separables states
\begin{equation}\label{lurs2}
\sum_{k=1}^n \Delta^2 (A_k\otimes \mathbb{I} + \mathbb{I}\otimes B_k) \ge C_A + C_B
\end{equation}
has to hold. Any quantum states which violate the equation (\ref{lurs2}) is entangled. The physical interpretation of (\ref{lurs2}) may be stated as follows:
separable states always inherit the uncertainty  relations which hold for their reduced states \cite{lur2}. The LURs criterion is strong and can be implemented 
with local measurements. Nevertheless, they have some disadvantages in the sense it is not clear which operators $A_k$ and $B_l$ one should choose to detect a given entangled state. Also, LURs can by construction characterize separable states only and they do not apply to other convex sets.  More about LURs criterion and its comparison with other criteria, for instance the CCN criterion and the criterion based on witness can be found in \cite{lur1, lur4, lur5}.

\item {\it The Li-Qiao criterion}

A recent criterion for the separability of an arbitrary bipartite mixed state is proposed by Li and Qiao \cite{liqiao} by virtue of the multiplicative Horn's problem \cite{horn1, horn2, horn3}. In the Li-Qiao's criterion, a complete and finite set of inequalities to determine the separability of compound system is obtained, which may be viewed as trade-off relations between the quantumness of subsystems.
The Li-Qiao's work follows the work initiated by Horodecki and al. \cite{horo} and uses the Bloch vector representation introduced to the separability problem by J. De Vicente \cite{vicente}.

Let's recall that a mixed bipartite state of particles $A$ and $B$ is separable if and only if it can be expressed as 
\begin{equation}\label{liqiao1}
\rho_{AB} = \sum_{i=1}^Lp_i\rho_{A,i}\otimes \rho_{B,i},
\end{equation}
where $p_i > 0$ with $\sum_{i=1}^L p_i = 1$, and $\rho_{A,i}, \: \rho_{B,i}$ are local density matrices of the particles $A$ and $B$. In the De Vicente criterion, we have seen that an arbitrary $d_A \times d_B$ dimensional bipartite state in the Bloch representation is
\begin{equation}\label{liqiao2}
\rho_{AB} = \frac{1}{d_Ad_B}
\mathbb{I}_{d_A}\otimes \mathbb{I}_{d_B} + \frac{1}{2d_B}\vec{r}\cdot\vec\lambda\otimes \mathbb{I}_{d_B} + \frac{1}{2d_A}\mathbb{I}_{d_A}\otimes \vec{s}\cdot \vec\sigma + \sum_{i = 1}^{d_A^2-1}\sum_{j=1}^{d_B^2 -1}\tau_{ij}\lambda_i \otimes \sigma_j.
\end{equation}
Since a bipartite state is separable if it can be decomposed as the sum of direct products of local density matrices as shown in equation (\ref{liqiao1}). The Li-Qiao necessary and sufficient condition for the separability of $\rho_{AB}$ in equation (\ref{liqiao2}) is given by 
\begin{equation}\label{liqiao3}
\sum_{i =1}^L p_i\vec{a}_i = \vec r; \quad \sum_{j=1}^L p_j\vec b_j = \vec s,\quad \sum_{k=1}^Lp_k\vec a_k {\overrightarrow{b}}^T_k = \tau,
\end{equation}
where $L$ stands for the number of local states needed in the separable decomposition, $p_i > 0$, $\sum_{i=1}^L p_i = 1$ and $\rho_i^A = \frac{1}{d_A}\mathbb{I} + \frac{1}{2}\vec{a}_i\cdot \vec{\lambda}$ and $\rho_i^B = \frac{1}{d_A}\mathbb{I} + \frac{1}{2}\vec{b}_i\cdot \vec{\sigma}$, with $\vec{a}_i, \; \vec{b}_j$ being the Block vectors of the decomposed local quantum states. The equation (\ref{liqiao3}) can be set in matrix form as follows.
\begin{equation}
M_a \vec{p} = \vec r;\quad M_b\vec{p} = \vec s; \quad M_{ap}M_{bp}^T = \tau.
\end{equation}
Here $M_a = (\vec a_1,\vec a_2,\ldots, \vec{a}_L)$, and $M_b = (\vec b_1,\vec b_2,\ldots, \vec{b}_L)$, with $\vec a_i$, $\vec b_j$ being $d_A^2 -1$, $d_B^2 -1$ 
dimensional real vectors respectively; $\vec p = (p_1, p_2, \ldots p_L)^T$, and $M_{ap} = M_aD_p^{\frac{1}{2}}$, $M_{bp} = M_bD_p^{\frac{1}{2}}$ with 
$D_p = \textrm{diag}\{p_1,p_2, \ldots, p_L\}$. 

Though in principle the Li-Qiao criterion is necessary and sufficient, its physical significance and pratical applications need to be exemplified. So, to the best of our knowledge, the Li and Qiao criterion is rather a reformulation of the definition of entanglement.

\end{enumerate}

\section{Conclusion}\label{sec4}

In this work, we have revisited the following criteria in bipartite systems focusing in the case of a system of two quantum bits: 1) Bell inequalities criterion, 2) Schmidt decomposition criterion, 3) the entropy of entanglement criterion, 4) the PPT or Horodecki-Peres criterion, 5) the reduction criterion, 6) the concurrence criterion, 7) the majoration criterion, 8) the computable cross norm or realignment (CCNR) criterion, 9) the correlation matrix or de Vicente criterion, 10) the positive maps criterion, 11) the criterion based on  entanglement witnesses, 12) the local uncertainty relations (LURs) criterion, 13) the Li-Qiao criterion, 14) the SIC POVMs criterion.
While we have listed positive maps and entanglement witness criteria in the ``non -operational " criteria, it is good drawing the attention on the fact that these two criteria are the basis for many of the operational criteria are nothing but examples of positive but not completely positive maps. As for entanglement witnesses, it is common in experiments to know the prepared state before hand and to use an entanglement witness tailored to the experiment. From this point of 
view, entanglement witnesses are more operational than any other criterion.

 We do not provide demonstrations of the theorems, propositions, nor discuss the comparition between the various criteria since these elements are in the references cited. We do not pay attention to the higher dimensional case. For instance, we do not include the criterion for entanglement detection based on covariance matrices for an arbitrary set of observables which is said to be suitable for higher dimensions \cite{covma}. This review provides a good starting point for reading about the topic and directs the interested reader to more in-depth resources.

All the criteria of separability are mathematically based, 
for instance the characterization and classification of positive maps on $C^\star$ algebras \cite{stro, jami, choi, woro}, the theory of positive maps \cite{pm1,pm2,pm3,pm4},  the multiplicative Horn's problem \cite{horn1,horn2, horn3}, the theory of majoration \cite{major}. The Bell's inequality is only a mathematical theorem and the relation between Bell's inequalities and convex geometry is also well-known \cite{geobell}. The use of uncertainty arguments to study entanglement is well known from continous variables system \cite{epr2, reid, duan}. 

The separability of quantum states is direclty linked to unsolved challenges of mathematics concerning linear algebra and geometry, functional analysis and, in particular, the theory of $C^\star$-algebra. For instance the Gel'fand, Naimark and Segal (GNS) construction of representation of $C^\star$- algebra of observables allows to obtain a representation space for the subalgebra such that its decomposition into irreductible subspaces can be used to study quantum correlations \cite{balachandran}. The distillability problem, that is the question when the state of a composite quantum system can be transformed to an entangled pure state using local operations, is an other problem that is related to challenging open questions of modern mathematics. That is why some
details of mathematical basics tools that would help the beginners to be less
frightened by the not basics mathematics of quantum entanglement \cite{lgouba}.

\end{document}